\documentclass[useAMS,usenatbib]{mn2e}
\usepackage{epsfig}
\usepackage{graphics}
\usepackage[usenames]{color}
\usepackage{endnotes}

\usepackage{amssymb}
\usepackage{amsmath}
\usepackage{times}
\usepackage{subfigure}
\newcommand{\be}{\begin{equation}}
\newcommand{\ee}{\end{equation}}
\newcommand{\bdm}{\begin{displaymath}}
\newcommand{\edm}{\end{displaymath}}
\newcommand{\bea}{\begin{eqnarray}}
\newcommand{\eea}{\end{eqnarray}}

\newcommand{\msun}{M_\odot}
\def\lsim{\lower.5ex\hbox{$\; \buildrel < \over \sim \;$}}

\title[GW background in the pulsar timing band]
{Systematic investigation of the expected gravitational wave signal from supermassive black hole binaries in the pulsar timing band}

\author[A. Sesana]
       {A. Sesana$^{1}$\thanks{E-mail: alberto.sesana@aei.mpg.de}
\\
$^{1}$ Max-Planck-Institut f{\"u}r Gravitationsphysik, Albert Einstein
Institut, Am Muhlenberg 1, 14476, Golm, Germany \\
}

\begin{document}

\date{}

\pagerange{\pageref{firstpage}--\pageref{lastpage}} \pubyear{2010}

\maketitle

\label{firstpage}

\begin{abstract}
In this letter we carry out the first systematic investigation of the expected gravitational wave (GW) background generated by supermassive black hole (SMBH) binaries in the nHz frequency band accessible to pulsar timing arrays (PTAs). We take from the literature several estimates of the redshift dependent galaxy mass function and of the fraction of close galaxy pairs to derive a wide range of galaxy merger rates. We then exploit empirical black hole-host relations to populate merging galaxies with SMBHs. The result of our procedure is a collection of a large number of
%$5832$ 
phenomenological SMBH binary merger rates consistent with current observational constraints on the galaxy assembly at $z<1.5$. For each merger rate we compute the associated GW signal, eventually producing a large set of 
%$5832$ 
estimates of the nHz GW background that we use to infer confidence intervals of its expected amplitude. When considering the most recent SMBH-host relations, accounting for ultra-massive black holes in brightest cluster galaxies, we find that the nominal $1\sigma$ interval of the expected GW signal is only a factor of 3-to-10 below current PTA limits, implying a non negligible chance of detection in the next few years. 
\end{abstract}

\begin{keywords}
black hole physics - gravitational waves - pulsars: general - galaxies: evolution 
\end{keywords}
%###############################################################################
%###############################################################################
%###############################################################################
%###############################################################################

\section{INTRODUCTION}
\label{sec:Introduction}

Precision timing of an array of millisecond pulsars (PTA) provides a unique opportunity to get the very first low-frequency gravitational wave (GW) detection. The European Pulsar Timing Array \cite[EPTA,~][]{ferdman10}, the Parkes Pulsar Timing Array \cite[PPTA,~][]{man12} and the North American Nanohertz Observatory for Gravitational Waves \cite[NANOGrav,~][]{jenet2009}, joining together in the International Pulsar Timing Array \cite[IPTA,~][]{hobbs2010}, are constantly improving their sensitivity in the frequency range of $\sim10^{-9}-10^{-6}$ Hz. 
%In the coming years, the Chinese five hundred meter aperture spherical telescope \cite[FAST,~][]{smits09} and the planned Square Kilometer Array \cite[SKA,~][]{lazio2009} will provide major leaps in sensitivity. 
Inspiralling supermassive black hole (SMBH) binaries populating merging galaxies throughout the Universe are expected to generate the dominant signal in this frequency band \cite[see, e.g.][]{rr95,jaffe03,wl03,sesana08}. 
Generally speaking, the expected amplitude of the signal depends on the pace at which SMBH binary mergers occur along cosmic history, and on their typical masses. Both quantities are poorly determined observationally, allowing for a wide range of GW signal amplitudes. Theoretical models of SMBH evolution within the standard hierarchical framework of galaxy formation indicate a typical GW strain amplitude $A\sim10^{-15}$ at $f=1/$yr \citep{wl03,sesana08,ravi12}, with an uncertainty of $\approx$0.5dex \citep{sesana08}. However, a recent investigation by \cite{mcwilliams12}, based on a phenomenological model in which the low redshift massive galaxy assembly is driven by mergers only, predicts a higher background with a fiducial amplitude $A\sim6\times10^{-15}$. Though useful, all the aforementioned models employ specific recipes for the galaxy assembly and/or for the growth of SMBHs, and a systematic investigation of the possible range of signals compatible with observational uncertainties is still missing. 
This is a particularly important issue to assess at this point for two reasons: (i) the best limit placed by PTAs on the GW background amplitude is $A=6\times10^{-15}$ \citep{vanh11}, close to theoretical predictions; (ii) SMBHs in brightest cluster galaxies (BCGs) were recently discovered to be more massive than expected \citep{fabian12}, resulting in a revision to the established SMBH-host relations \citep{mcconnell12} that might push the expected GW background level closer to current upper limits, implying possible detection in the next few years.

As part of the common effort of the EPTA collaboration to detect GWs with pulsar timing \citep{vanh11}, we present here the first {\it systematic investigation} of the range of GW signal amplitudes consistent with {\it observationally based} estimates of the SMBH assembly in the low redshift Universe. The manuscript is organized as follows. In Section 2 we describe our model for generating the GW background and we test it against a range of observational constrains. We present and discuss our main results in Section 3, and draw our conclusions in Section 4. Throughout the paper we assume a concordance $\Lambda$--CDM universe with $\Omega_M=0.27$, $\Omega_\lambda=0.73$ and $h=0.7$. Unless otherwise specified, we use geometric units where $G=c=1$.

\section{Building the GW background from astrophysical observables}
\label{subsec:model}

\subsection{Mathematical description of the GW background}
Consider a cosmological population of merging SMBH binaries. Each merging pair is characterized by the masses of the two holes $M_{\bullet,1}>M_{\bullet,2}$, defining the mass ratio $q_\bullet=M_{\bullet,2}/M_{\bullet,1}${\footnote{to avoid confusion, we denote the masses and mass ratio of the SMBH binary as $M_{\bullet,1},M_{\bullet,2},q_\bullet$, whereas plain $M$ and $q$ are used for galaxies}}. Following \cite{sesana08} \citep[see also][]{phinney01}, the characteristic amplitude $h_c$ of the GW signal generated by such population is given by
%%%%%%%%%%%%%%%%%%%%%%%%%%%%%%%%%
\begin{equation}
h_c^2(f) =\frac{4}{\pi f^2}\int \int \int 
dzdM_{\bullet,1}dq_\bullet \, \frac{d^3n}{dzdM_{\bullet,1}dq_\bullet,}
{1\over{1+z}}~{{dE_{\rm gw}({\cal M})} \over {d\ln{f_r}}}\,.
\label{hcdE}
\end{equation}
%%%%%%%%%%%%%%%%%%%%%%%%%%%%%%%%%
Here, the energy emitted per logarithmic frequency interval is \citep{thorne87}
%%%%%%%%%%%%%%%%%%%%%%%%%%%%%%%%%
\begin{equation}
\frac{dE_{\rm gw}}{d\ln{f_r}}=\frac{\pi^{2/3}}{3}{\cal M}^{5/3}f_r^{2/3}\,,
\label{dedlnf}
\end{equation}
%%%%%%%%%%%%%%%%%%%%%%%%%%%%%%%%%
where we assumed circular binaries driven by GW emission only{\footnote{We keep this assumption throughout the paper. Eccentricity, together with other physical mechanisms (stellar scattering, gas torques) driving the binaries can modify the form of $dE_{\rm gw}/d\ln{f_r}$ given by equation (\ref{dedlnf}). We defer the investigation of these issues to future work.}}, ${\cal M}=(M_{\bullet,1}M_{\bullet,2})^{3/5}/(M_{\bullet,1}+M_{\bullet,2})^{1/5}$ is the chirp mass of the binary and $f_r=(1+z)f=$ is the GW rest frame frequency, which is twice the binary Keplerian frequency. The quantity $d^3n/(dzdM_{\bullet,1}dq_\bullet)$ represents the differential merger rate density (i.e. number of mergers per comoving volume) of SMBH binaries per unit redshift, mass and mass ratio. For convenience, we decided to keep it a function of $M_{\bullet,1}$ and $q_\bullet$ instead of ${\cal M}$ only.

It is straightforward to show \citep{phinney01} that the predicted characteristic amplitude scales as $\propto f^{-2/3}$, with a normalisation that depends on the details of the merging binary population, and is usually represented as
\citep[see, e.g.,][]{jen06}:
% So in general, one can find in literature (e.g. Jenet et al. 2006):
%%%%%%%%%%%%%%%%%%%%%%%%%%%%%%%%%
\begin{equation}
h_c(f) =A\, \left(\frac{f}{{\rm yr}^{-1}}\right)^{-2/3}\,,
\label{hcpar}
\end{equation}
%%%%%%%%%%%%%%%%%%%%%%%%%%%%%%%%%
where $A$ is a model dependent constant that represents the amplitude of the signal at the reference frequency $f=1 {\rm yr}^{-1}$. Since observational limits on the GW background are usually given in terms of $A$ \citep{jen06,vanh11,demorest12}, we keep the parametrization given by equation (\ref{hcpar}) in this letter, and we investigate the range of $A$ predicted by phenomenological models of the SMBH assembly based on observations.

Before proceeding, it is worth mentioning that the shortcomings of equation (\ref{hcdE}) in catching the relevant features of the GW signal emitted by a realistic population of {\it quasi monochromatic sources} were extensively investigated by \citep{sesana08,sesana09,ravi12}. Although equation (\ref{hcdE}) fails in describing small number statistics effects and the intrinsic non Gaussianity of the signal, it is sufficient to describe its expected overall amplitude, which is our main interest here. In a companion paper (Sesana in preparation) we will carry out a more systematic study of all the relevant signal features, including statistics of expected individually resolvable sources.

\subsection{Determination of the SMBH binary merger rate}
Since the energy emitted per logarithmic frequency interval is fixed by General Relativity (in the approximation of circular GW driven binaries), the typical background strength $A$ depends on the SMBH binary differential merger rate only. In contrast to our past work \citep{sesana08,sesana09}, we take here an observational approach to determine $d^3n/dzdM_{\bullet,1}dq_\bullet$. We proceed in two steps: (i) we determine {\it from observations} the {\it galaxy} merger rate $d^3n_G/dzdMdq$ (in a merging galaxy pair, $M$ and $q<1$ are the mass of the primary galaxy and the mass ratio respectively), and (ii) we populate merging galaxies with SMBHs according to empirical black hole mass--galaxy host relations found in the literature.

\subsubsection{Galaxy merger rate}
The galaxy differential merger rate can be written as
\begin{equation}
\frac{d^3n_G}{dzdMdq}=\frac{\phi(M,z)}{M\ln{10}}\frac{{\cal F}(z,M,q)}{\tau(z,M,q)}\frac{dt_r}{dz}.
\label{galmrate}
\end{equation}
Here, $\phi(M,z)=(dn/d{\rm log}M)_z$ is the galaxy mass function measured at redshift $z$; ${\cal F}(M,q,z)=(df/dq)_{M,z}$ is the differential fraction of galaxies with mass $M$ at redshift $z$ paired with a secondary galaxy having a mass ratio in the range $q, q+\delta{q}$; $\tau(z,M,q)$ is the typical merger timescale for a galaxy pair with a given $M$ and $q$ at a given $z$; and $dt_r/dz$ converts a proper time rate into a redshift rate and is given by standard cosmology. The reason for writing equation (\ref{galmrate}) is that $\phi$ and ${\cal F}$ can be directly measured from observations, whereas $\tau$ can be inferred by detailed numerical simulations of galaxy mergers, as discussed below.

We take three different galaxy stellar mass functions from the literature \citep{borch06,drory09,ilbert10} and match them with the local mass function \citep{bell03}, to obtain three fiducial $\phi_z(M)$. To each fiducial mass function we add an upper and a lower limit accounting for the errors given by the authors on the function best fit parameters, plus an additional 0.1dex systematic error due to uncertainties in the determination of the galaxy masses, for a total of 9 galaxy mass functions. For all mass functions we separate early type and late type galaxies. We restrict our calculation to $z<1.3$ and $M>10^{10}\msun$, since these are the systems contributing the largest fraction of the GW signal. By extrapolating our calculations 
%to higher redshifts and lower masses, 
we found that merging pairs residing in galaxies with $M<10^{10}\msun$ or at $z>1.3$ can contribute at most $5\%$ to the signal amplitude, and can be safely neglected.

We consider four studies of the evolution of the galaxy pair fraction \citep{bundy09,deravel09,lopez12,xu12}. Pair fractions are usually integrated over some range of $q$ and are given in different mass bins in the form $f(z)=f_0(1+z)^{\gamma}$. A good proxy for the observed pair $q$ distribution is $df/dq\propto q^{-1}$. Given $f(z)$, we can therefore simply write $df/dq(z)=-f(z)/(q{\ln q_{m}})$, where $q_m$ is the minimum mass ratio selected in counting pairs. Each author applies different criteria as for $q_m$, mass and redshift range, and the maximum projected distance $d_{\rm max}$ below which two galaxies are considered a bound pair, as detailed in table \ref{tabpair}. Also in this case, for each of the 4 fiducial models we consider an upper and a lower limit taking into account the the errors in the best fit parameters $f_0, \gamma$, as reported by the authors, to get a total of 12 pair fraction models. When necessary, we extrapolate the pair fraction estimates to cover the full mass and redshift range of interest ($z<1.3$ and $M>10^{10}\msun$). When pair counting for different galaxy types are available, we apply them to the corresponding galaxy type mass function, otherwise, we assume the same pair fraction for early and late type galaxies. \cite{lopez12} also provide pair fractions for 'minor mergers', i.e., for $0.25>q>0.1$. We checked that including those in our calculation enhances the background by a factor 0.06dex ($\lesssim 15\%$) at most.

%%%%%%%%%%%%%%%%%%%%%%%%%%
\begin{table}
\begin{center}
\begin{tabular}{ccccc}
\hline
Paper & $q_m$ & $M_{\rm min}[\msun]$ & $d_{\rm max}$[kpc] & gal. type\\
\hline
Bundy et al. 2009    & 0.25      & $10^{10}$  & 20 & yes\\
de Ravel et al. 2009 & 0.25      & $10^{9.5}$ & 100 & no\\
Lopez et al. 2012    & 0.25      & $10^{11}$  & 30 & yes\\
Xu et al. 2012       & 0.4       & $10^{9.4}$ & 20 & no\\
\hline
\end{tabular}
\end{center}
\caption{Overview of the pair fraction selection performed in the paper used in this work. See text for details.}
\label{tabpair}
\end{table}
%%%%%%%%%%%%%%%%%%%%%%%%%%%%%%%%%%%%%%%

Galaxy merger timescales ${\tau}$ were carefully estimated by \cite{kit08} using mock catalogues of galaxy pairs in the Millennium simulation \citep{springel05}. In their equation (10) they provide the average merger timescale as a function of $M$, $z$, and projected distance $d_p$. We complemented their equation (10) with a $\approx q^{-0.3}$ dependence extracted by fitting the results of a set of full hydrodynamical simulation of galaxy mergers presented by \cite{lotz10}. In doing this, we noticed that the merger timescales given by \cite{lotz10} are a factor of two shorter than those given by \cite{kit08}; we therefore adopted two different normalizations to get a 'fast' and a 'slow' merger scenario.
  
We interpolate all the measured $\phi,{\cal F},\tau$ on a fine 3-D grid in $(z,M,q)$, to numerically obtain $9\times12\times2=216$ differential galaxy merger rates. Note that typical values of $\tau$ are of the order of a Gyrs, therefore, the merger rate at a given $(z,M,q)$ point in the grid is obtained by evaluating $\phi$ and ${\cal F}$ at $(z+\delta{z},M,q)$, where $\delta{z}$ is the redshift delay corresponding to the merging time $\tau$. Note that by doing this, we implicitly assume that all SMBH binaries coalesce instantaneously at the merger time of their hosts.
%istantaneous SMBH binary coalescence as the pair of galaxies merges.

\subsubsection{Black hole-host relations}
We assign to each merging galaxy pair SMBHs with masses drawn from 9 different SMBH-galaxy relations found in the literature (see table \ref{tabrel}). We write them in the form
\begin{equation}
{\rm log}_{10} M_{\bullet}=\alpha+\beta{\rm log}_{10}X,
\label{scalingrel}
\end{equation}
where $X=\{\sigma/200$km s$^{-1}$, $L_i/10^{11}L_{i,\sun}$ or $M_{\rm bulge}/10^{11}\msun\}$, being $\sigma$ the stellar velocity dispersion of the galaxy bulge, $L_i$ its mid-infrared luminosity, and $M_{\rm bulge}$ its stellar mass. Each relation is characterized by an intrinsic scatter $\epsilon$. $\alpha, \beta, \epsilon$ are listed in table \ref{tabrel}. The relations link $M_{\bullet}$ to the {\it bulge} properties, whereas our galaxy merger rates are function of the total stellar mass. We derive the bulge mass of each galaxy by multiplying the total stellar mass by a factor $f_{\rm bulge}$. We assume $f_{\rm bulge}=1$ for all early type galaxies with $M>10^{11}\msun$, declining to $f_{\rm bulge}=0.5$ at $M=10^{10}\msun$, whereas we assign a random $f_{\rm bulge}$ in the range 0.1-0.3 to late type (i.e., disk  dominated) galaxies. Although this prescription is somewhat arbitrary, we found that the results are almost independent of $f_{\rm bulge}$ as long as massive early type galaxies retain a bulge fraction of order unity, which is a well established observational fact. Given the bulge mass, we estimate $L_i$ by inverting the $M_{\rm bulge}-L_i$ relation given by \cite{sani11}, and we compute $\sigma$ by fitting a broken power-law to the $z=0$ $\sigma-M_{\rm bulge}$ data presented by \cite{robertson06}. When converting $M_{\rm bulge}$ into $\sigma$ we apply a multiplication factor $(1+z)^{0.3}$, to account for the observational fact that galaxies of a given mass at higher redshift are more concentrated and have larger velocity dispersions than galaxies of the same mass at lower redshift \citep[see, e.g.][]{lopez12}. Having derived $M_{\rm bulge}, L_i$ and $\sigma$, we can populate galaxies with SMBHs. We apply a further $(1+z)^{0.3}$ correction to the scaling relations involving $M_{\rm bulge}$ and $L_i$. This redshift dependence improves the match of the SMBH mass density ($\rho_{\rm BH}$) redshift evolution given by our models with other estimates found the literature (see figure \ref{fig2}), and we checked that our results are basically independent of it.

%%%%%%%%%%%%%%%%%%%%%%%%%%
\begin{table}
\begin{center}
\begin{tabular}{ccccc}
\hline
Paper & $X$ & $\alpha$ & $\beta$ & $\epsilon$\\
\hline
\cite{haring04}     & $M_{\rm bulge}$ & 8.2 & 1.12 & 0.30\\
\cite{sani11}       & $M_{\rm bulge}$ & 8.2 & 0.79 & 0.37\\
\cite{beifiori12}    & $M_{\rm bulge}$ & 7.84 & 0.91 & 0.46\\
\cite{mcconnell12}    & $M_{\rm bulge}$ & 8.46 & 1.05 & 0.34\\
\cite{sani11}        & $L_i$         & 8.19 & 0.93 & 0.38\\
\cite{gultekin09}    & $\sigma$      & 8.23 & 3.96 & 0.31\\
\cite{graham11}      & $\sigma$      & 8.13 & 5.13 & 0.32\\
\cite{beifiori12}   & $\sigma$      & 7.99 & 4.42 & 0.33\\
\cite{mcconnell12}  & $\sigma$      & 8.33 & 5.57 & 0.40\\

\hline
\end{tabular}
\end{center}
\caption{List of parameters $\alpha$, $\beta$ and $\epsilon$. See text for details.}
\label{tabrel}
\end{table}
%%%%%%%%%%%%%%%%%%%%%%%%%%%%%%%%%%%%%%%
%%%%%%%%%%%%%%%%%%%%%%%%%%%%%%%%%%%%%%%%%
\begin{figure}
\begin{tabular}{c}
\includegraphics[scale=0.35,clip=true,angle=0]{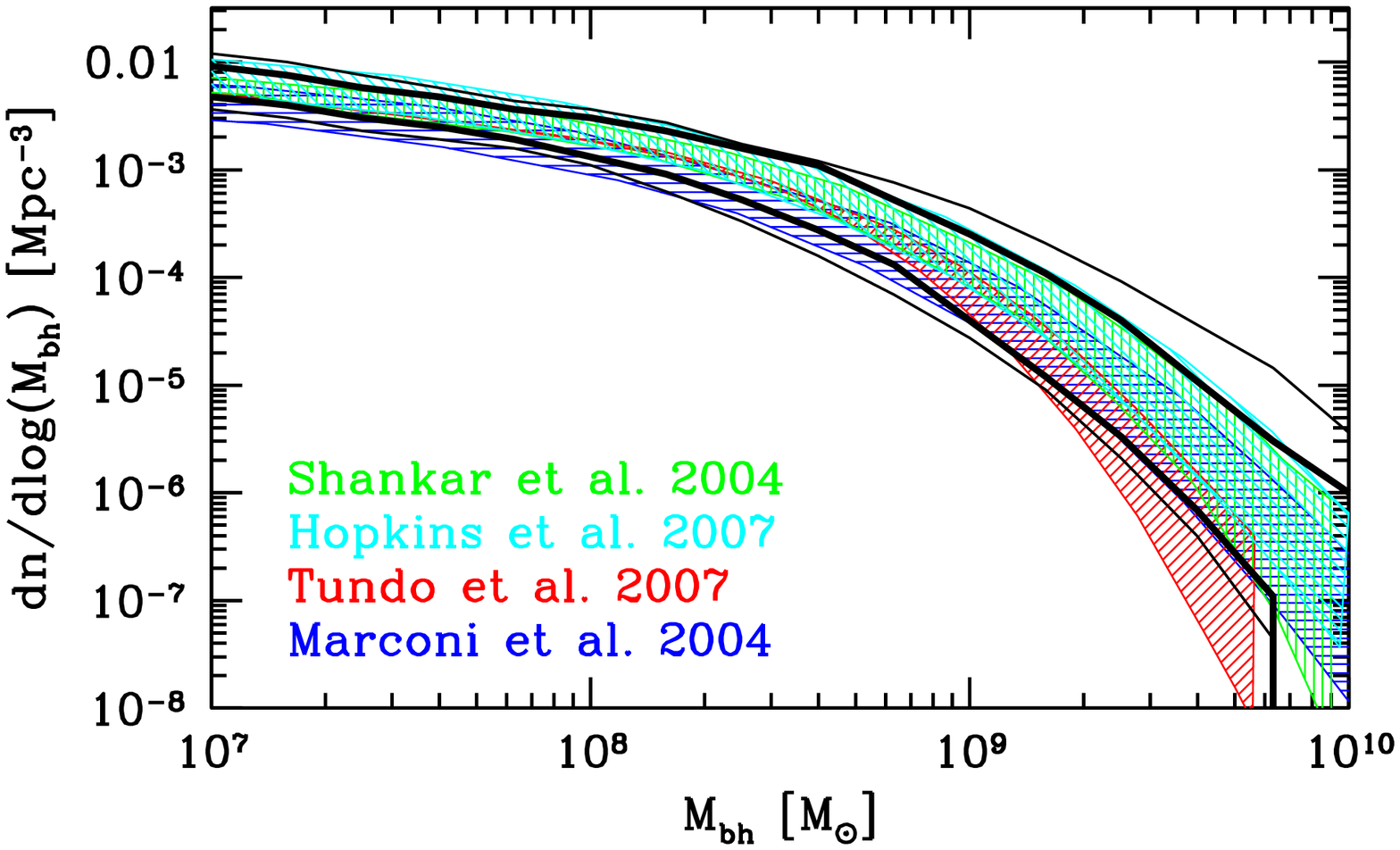}\\
\includegraphics[scale=0.35,clip=true,angle=0]{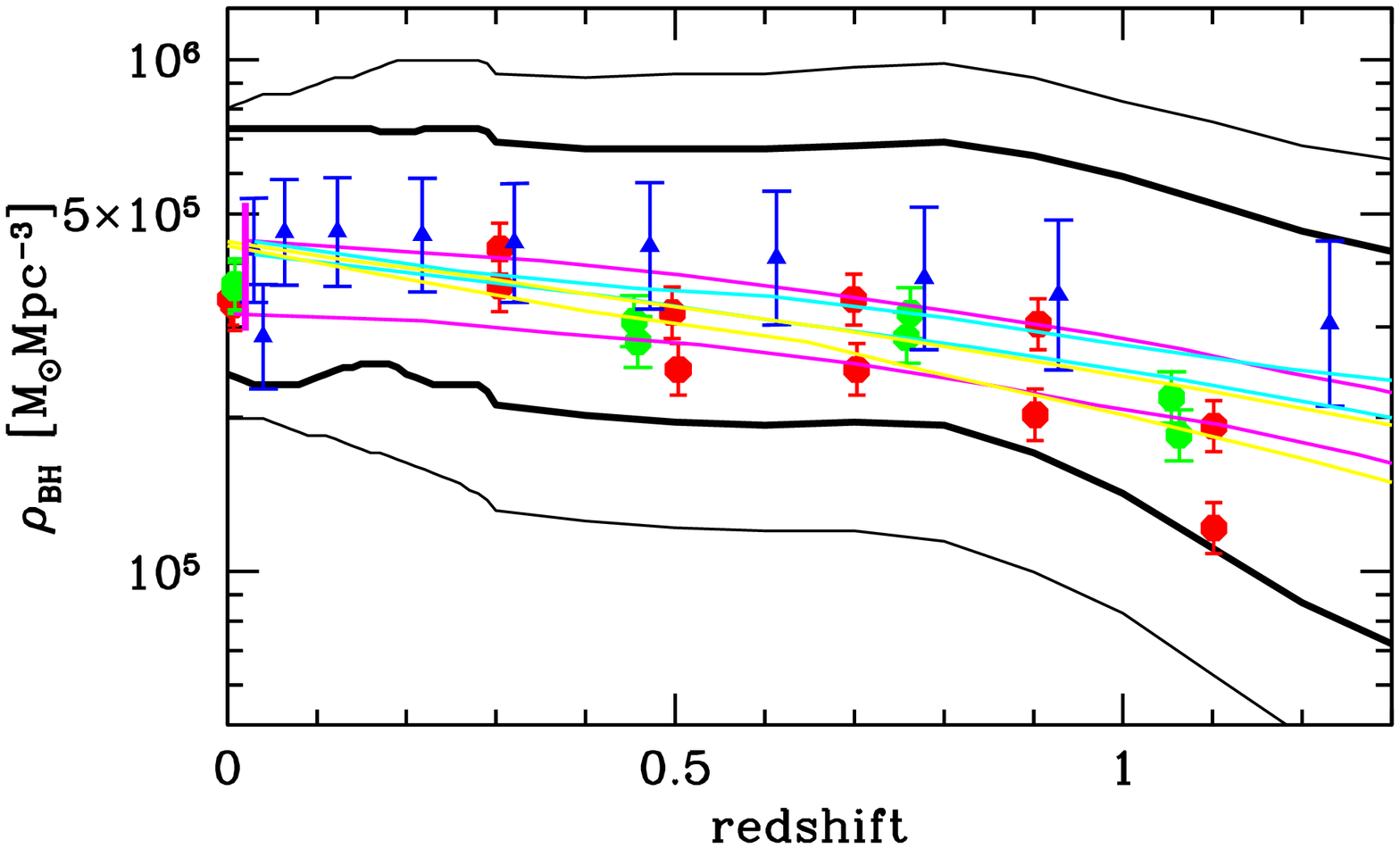}\\
\end{tabular}
\caption{Upper plot: local SMBH mass function. Thick and thin solid black lines enclose the areas corresponding to $68\%$ and $95\%$ confidence levels given by our models. Colored shaded areas are SMBH mass functions estimated by \protect\cite{marconi04,shankar04,hopkins07,tundo07}. Lower plot: redshift evolution of the total SMBH mass density. Thick and thin solid black lines have the same meaning as in the upper plot. Red and green dots are from \protect\cite{zhang12}, blue dots are from \protect\cite{hopkins07}, cyan and yellow lines are from \protect\cite{merloni08}, magenta lines bracket the $1\sigma$ uncertainty given by \protect\cite{shankar04}, and the thick magenta line is the estimated uncertainty range at $z=0$ from \protect\cite{shankar09}.}
\label{fig1}
\end{figure}
%%%%%%%%%%%%%%%%%%%%%%%%%%%%%%%%%%%%%%%%%

We assign to each merger remnant a total bulge mass equal to the sum of the total stellar masses of the merging systems, i.e., $f_{\rm bulge}=1$. Furthermore, we correlate the masses of the merging SMBHs either to the properties of the two merging galaxies or to those of the merger remnant, following the scheme described in Section 2.2 of \cite{sesana09}. This gives us three slightly different mass estimations for the SMBHs forming the binary for each adopted scaling relation.

We combine the $9\times3=27$ different ways to populate the merging galaxies with SMBHs together with the 216 galaxy merger rates to obtain 5832 different SMBH binary merger rates $d^3n/dzdM_{\bullet,1}dq_\bullet$, consistent with current observations of the evolution of the galaxy mass function and pair fractions at $z<1.3$ and $M>10^{10}\msun$ and with the empirical SMBH-host relations published in the literature. We give equal credit to each model, and we generate 5832 GW signals, sufficient to place reasonable confidence levels for the expected amplitude according to {\it current observational constraints}. Our approach is modular in nature, and it is straightforward to expand the range of model to include new estimates of all the quantities involved.

\subsection{Validation of the models}
%%%%%%%%%%%%%%%%%%%%%%%%%%%%%%%%%%%%%%%%%
%\begin{figure}
%  \hspace{-0.0cm}
%  %\includegraphics[width=1.15\linewidth]{plots/FIG_ECC_M.jpg}
%  \includegraphics[width=1.0\linewidth]{fig_mf.ps}
%  \caption{Local SMBH mass function. Thick and thin solid black lines enclose the areas corresponding to $68\%$ and $95\%$ confidence levels given by our models. Colored shaded areas are SMBH mass function estimated by \protect\cite{marconi04,shankar04,hopkins07,tundo07}.} 
% \label{fig1}
%\end{figure}
%%%%%%%%%%%%%%%%%%%%%%%%%%%%%%%%%%%%%%%%%
%%%%%%%%%%%%%%%%%%%%%%%%%%%%%%%%%%%%%%%%%
%\begin{figure}
%  \hspace{-0.0cm}
%  %\includegraphics[width=1.15\linewidth]{plots/FIG_ECC_M.jpg}
%  \includegraphics[width=1.0\linewidth]{fig_rho.ps}
%  \caption{Redshift evolution of the total SMBH mass density. Thick and thin solid black lines enclose the areas corresponding to $68\%$ and $95\%$ confidence levels given by our models. Red and green dots are from \protect\cite{zhang12}, blue dots are from \protect\cite{hopkins07}, cyan and yellow lines are from \protect\cite{merloni08}, magenta lines bracket the $1\sigma$ uncertainty given by \protect\cite{shankar04}, and the thick magenta line is the estimated uncertainty range at $z=0$ from \protect\cite{shankar09}.}
% \label{fig2}
%\end{figure}
%%%%%%%%%%%%%%%%%%%%%%%%%%%%%%%%%%%%%%%%%
Although the evolution of the SMBH masses is not followed self--consistently in our models, in figure \ref{fig1} we validate them by comparing the local SMBH mass function and the redshift evolution of the total SMBH density with several estimates found in the literature. We also checked that the predicted range of galaxy and SMBH merger rates as a function of mass and redshift are broadly consistent (though with a large scatter) with those derived from our previous models constructed on top of the Millennium Simulation \citep{sesana09} or exploiting semianalytical merger trees \citep{sesana08}. In the latter approach we evolve the SMBH population self--consistently. In figure \ref{fig1} we show the nominal $1\sigma$ and $2\sigma$ confidence levels (i.e. the range in which $68\%$ and $95\%$ of our models are contained) of the estimated local SMBH mass function and mass density as a function of $z$. The agreement with independent results published in the literature is excellent. We notice that we allow for slightly larger values of both quantities with respect to published results. This is because the \cite{mcconnell12} scaling relations, that include the recently measured ultra-massive SMBHs in BCGs, predict SMBH masses which are 0.2-to-0.4dex larger than previous estimates at the high mass end. Those models will result in larger amplitude of the GW signal, which might be soon directly tested with PTA observations.

%###############################################################################
\section{Results}
\label{results}
%%%%%%%%%%%%%%%%%%%%%%%%%%%%%%%%%%%%%%%%%
\begin{figure}
\includegraphics[scale=0.4,clip=true,angle=0]{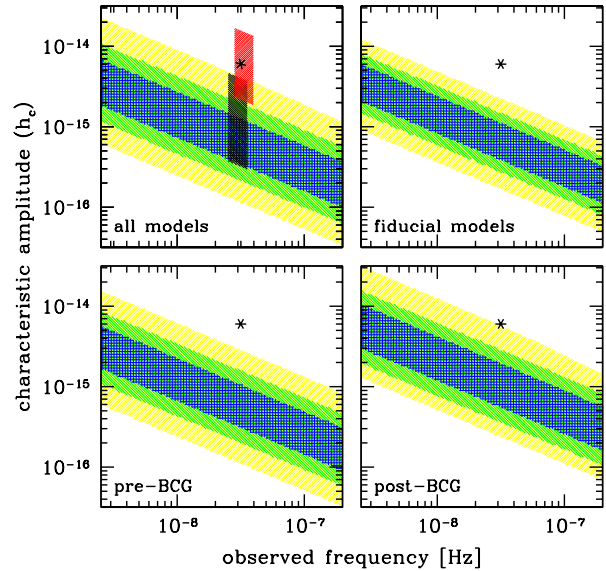}
\caption{Characteristic amplitude of the GW signal. Shaded areas represent the  $68\%$, $95\%$ and $99.7\%$ (nominally $1\sigma,2\sigma,3\sigma$) confidence levels given by our models. In each panel, the black asterisk marks the best current limit from \protect\cite{vanh11}. Shaded areas in the upper left panel refer to the $95\%$ confidence level given by \protect\cite{mcwilliams12} (red) and the uncertainty range estimated by \protect\cite{sesana08}. See text for discussion.}
\label{fig2}
\end{figure}
%%%%%%%%%%%%%%%%%%%%%%%%%%%%%%%%%%%%%%%%%
%%%%%%%%%%%%%%%%%%%%%%%%%%%%%%%%%%%%%%%%%
\begin{figure}
\includegraphics[scale=0.4,clip=true,angle=0]{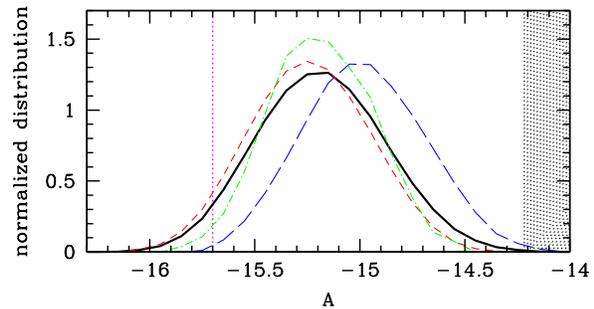}
\caption{Normalized distributions of the expected GW amplitude $A$ at $f=1$yr$^{-1}$. Black solid line, all models; green dot--dashed line, fiducial models only; red short--dashed line, models antecedent SMBH measurements in BCGs; blue long--dashed, models including SMBH measurements in BCGs. The shaded area marks the region excluded by current PTA limits, whereas the solid dotted line represent what can be achieved by timing 20 pulsars at 100ns rms precision for 10 years.}
\label{fig3}
\end{figure}
%%%%%%%%%%%%%%%%%%%%%%%%%%%%%%%%%%%%%%%%%

Our main result is shown in figure \ref{fig2}, where we plot confidence levels on the GW characteristic amplitude given by our models. When considering the whole set of models (upper left panel), the $68\%$ confidence region lies in the range $3.3\times10^{-16}<A<1.3\times10^{-15}$, corresponding to a factor of 4 uncertainty in the GW signal. The $99.7\%$ region extends much further, in the range $1.1\times10^{-16}<A<4.2\times10^{-15}$, corresponding to a factor $\approx 40$ uncertainty. Note that this latter upper bound is only a factor 1.5 below the best limit placed by \cite{vanh11}. Our 'democratic' approach to the problem gives the same weight to all the models. One can argue that models featuring the best estimates of the galaxy mass function and pair counts, should be considered more robust than those constructed using the upper or lower limits for the same quantities (see Section 2.2.1). If we restrict to 'fiducial models only', the scatter is mildly reduced, and the $68\%$ and $99.7\%$ confidence levels are set in the range $3.8\times10^{-16}<A<1.1\times10^{-15}$ and $1.7\times10^{-16}<A<2.2\times10^{-15}$ respectively (upper right panel). Things become much more interesting if we consider only the SMBH-host relations updated to include the recent measurements of ultra-massive black hole in BCGs \citep{mcconnell12}. As expected, the signal is boosted-up, bringing the $68\%$ and $99.7\%$ confidence intervals to $5.6\times10^{-16}<A<2.0\times10^{-15}$ and $2.4\times10^{-16}<A<5.7\times10^{-15}$ respectively (lower right panel), a factor $\approx2$ larger then models featuring previous estimates of the SMBH-host relations (lower left panel). Although obtained with a completely different procedure, our confidence intervals are generally consistent with the estimated signal range given by \citep{sesana08}, whereas recent results by \cite{mcwilliams12} are marginally consistent (at a $3\sigma$ level) with our findings. This is not surprising, since their purely merger driven SMBH evolution naturally produces the highest possible signal for a given SMBH mass function. Within a year, IPTA observations will therefore be able to test the whole amplitude range predicted by this scenario (van Haasteren, private communication) . In figure \ref{fig3}, we plot the normalized distributions of $A$ given by all our models. The overall distribution (solid--black line) has a neat gaussian shape, in agreement with the central limit theorem. This is the sign that none of the ingredients of our model (galaxy mass function and pair fraction, coalescence time, SMBH-host relation) plays a particularly dominant role in determining the signal amplitude. The shaded area, marking the region of $A$ excluded by current limits, already overlaps with the long tails of the distributions. Roughly speaking, the maximum $A$ detectable by a PTA with a signal-to-noise ratio of 5 is given by \citep{sesana08} 
\begin{equation}
A\approx8\times10^{-16}\frac{\delta{t}_{\rm rms}}{100\,{\rm ns}}\left(\frac{N_r}{100}\right)^{-1/4}\left(\frac{N_p}{20}\right)^{-1/2}\left(\frac{T_{\rm obs}}{5\,{\rm yr}}\right)^{-5/3},
\end{equation}
where $\delta{t}_{\rm rms}$ is the rms residual of each individual measurement (assumed to be the same for each pulsar), $N_r$ is the number of measured residuals for each pulsar, $N_p$ is the number of pulsars in the array and $T_{\rm obs}$ is the duration of the experiment. Observations of 20 pulsars at 100ns rms precision for 10 years will allow to detect a signal of $A\approx2\times10^{-16}$, which encompasses more than $95\%$ of the models presented here (dotted vertical line in figure \ref{fig3}).

\section{Conclusions}
\label{sec:conclusions}
We presented the first systematic investigation of the GW background generated by a cosmological population of SMBH binaries in the nHz frequency band, relevant to PTAs. We generated a grand total of 5832 SMBH binary merger rates, by coupling several observed galaxy mass functions and pair counts to phenomenological SMBH-host relations, and assuming merger timescale prescriptions derived by detailed hydrodynamical simulations of galaxy mergers. By construction, our models are consistent with observational constraints, and produce SMBH mass functions and mass density evolutions with redshift consistent with several independent estimates found in the literature. When considering all models, we find characteristic amplitudes of the signal at a frequency $1$yr$^{-1}$  in the range $3.3\times10^{-16}<A<1.3\times10^{-15}$ at $68\%$ confidence. However, recent measurement of ultra-massive black holes in BCGs led to a revision of the SMBH-host relations, rising their high-mass end by 0.2--0.4dex. Models based on these latter results predict amplitudes in the range $5.6\times10^{-16}<A<2.0\times10^{-15}$ and $2.4\times10^{-16}<A<5.7\times10^{-15}$ at $68\%$ and $99.7\%$ confidence respectively. Our results are broadly consistent with our previous work \citep{sesana08} and marginally consistent (at a $3\sigma$ level) with recent work by \cite{mcwilliams12}. We predict a nominal $3\sigma$ upper limit to the signal close to current limits placed by PTAs. If our models are correct, with an improvement of a factor of only three on current limits, there is a non negligible chance to make the first ever direct GW detection. Even a negative result will nevertheless allow us to constrain the assembly of the most massive galaxies at low redshift and how do they correlate with their hosts, turning PTAs into useful astrophysical probes. Looking further ahead, the timing of 20 pulsars at 100ns rms precision for 10 years (considered a feasible long term goal with current PTAs) is almost guaranteed to detect the GW background. The first direct GW detection, might not be so far off in the future.
   
\section{Acknowledgments}
A.S. acknowledges the support of the colleagues in the EPTA and of the DFG grant SFB/TR 7 Gravitational Wave Astronomy and by DLR (Deutsches Zentrum fur Luft- und Raumfahrt). 

\bibliographystyle{mn2e}
\bibliography{references}

\bsp

\label{lastpage}

\end{document}